\mag=\magstephalf
\mag=\magstep1
\pageno=1
\input amstex
\documentstyle{amsppt}
\TagsOnRight

\pagewidth{16.5 truecm}
\pageheight{23.0 truecm}
\vcorrection{-1.0cm}
\hcorrection{-0.5cm}
\nologo

\NoBlackBoxes
\font\twobf=cmbx12

\define \bx{\overline{x}}
\define \bbf{\overline{f}}

\define \blambda{\overline{\lambda}}

\define \ord{\roman{ord}}

\define \IV{ {\text{IV}} }
\define \X{ {\text{X}} }

\define \ft{{u}}
\define\hw{\widehat{\omega}}
\define\w{\omega}
\define \tvskip{\vskip 1.0 cm}
\define\ce#1{\lceil#1\rceil}
\define\dg#1{(d^{\circ}\geq#1)}
\define\Dg#1#2{(d^{\circ}(#1)\geq#2)}
\define\dint{\dsize\int}
\def\fp{\flushpar}

\define\s#1{\sigma_{#1}}
\define\tp#1{\negthinspace\left.\ ^t#1\right.}
\define\mrm#1{\text{\rm#1}}
\define\lr#1{^{\sssize\left(#1\right)}}
\redefine\ord{\text{ord}}

\redefine\qed{\hbox{\vrule height6pt width3pt depth0pt}}

{\centerline{\bf{Hyperelliptic Solutions of KdV and
KP equations: }}}
{\centerline{\bf{Reevaluation of Baker's Study on
Hyperelliptic Sigma Functions}}}

\author
Shigeki MATSUTANI
\endauthor
\affil
8-21-1 Higashi-Linkan Sagamihara 228-0811 Japan
\endaffil \endtopmatter


\centerline{\twobf Abstract }\tvskip

Explicit function forms of hyperelliptic solutions of Korteweg-de
Vries (KdV) and \break Kadomtsev-Petviashvili (KP)
equations were constructed
for a given curve $y^2 = f(x)$ whose genus is three.
This study was based upon the fact that about one hundred years ago
(Acta Math. (1903) {\bf{27}}, 135-156),
H.~F.~Baker essentially derived  KdV hierarchy and
KP equation by using bilinear differential operator ${\bold{D}}$,
identities of  Pfaffians, symmetric functions,
hyperelliptic $\sigma$-function and
$\wp$-functions; $\wp_{\mu \nu} =
 -\partial_\mu \partial_\nu \log \sigma$
$= - ({\bold{D}}_\mu {\bold{D}}_\nu \sigma \sigma)/2\sigma^2$.
The connection between his theory and the modern soliton theory
was also discussed.


\document

\vskip 1 cm
{\centerline{\bf{\S 1. Introduction}}}
\vskip 0.5 cm

In this article we will construct
explicit function forms of hyperelliptic solutions of
Korteweg-de Vries (KdV) and
Kadomtsev-Petviashvili (KP) equations
for a given curve $y^2 = f(x)$ whose genus is three,
along the lines of the study of H.~F.~Baker's sigma function
[B1, B2, B3].
This construction means  reevaluation of Baker's studies on
hyperelliptic
functions which were done one hundred years ago as a special
case of his studies of algebraic functions over  a general
compact Riemannian surface [B3].
Although his general theory has been already
known as the studies related to Baker-Akhiezer functions [B1, K1, K2],
the paper [B3]  published in 1903 might have been
left behind.

According to [B3], around 1898
he discovered series of partial differential equations
which lead hyperelliptic sigma function, $\sigma$,
and $\wp$-functions,
$\wp_{\mu\nu}
:= \partial_\mu \partial_\nu \log \sigma$.
If one saw the partial differential equations, he would
know that they are related to soliton equations
such as the KdV equations or the KP equations. However
 Baker's definition of parameters is twisted from
those in modern soliton theory. Further as the paper [B3]
requires knowledge of hyperelliptic $\sigma$  and $\wp$
functions which might not be familiar nowadays [B1, B2, \^O2],
it is not easy to understand its contents and to confirm
the derivation.
In this paper, we will give  correspondences
between his differential equations
and, the KP equation and first and second
equations of the KdV hierarchy in order to construct
explicit function forms of their periodic multi-soliton
solutions.

The identification between Baker's differential equations
and these soliton equations means that
Baker essentially discovered the KdV hierarchy and
the KP equation one hundred years ago.
In the study, he used the Pfaffian, symmetric functions,
bilinear operator ${\bold{D}}$,
hyperelliptic sigma function $\sigma$ and $\wp$-functions;
$\wp_{\mu,\nu}= -
({\bold{D}}_\mu {\bold{D}}_\nu \sigma \sigma)/2\sigma^2 $.

In this paper, we will comment on
its relation to soliton theory in Sec 4.
As we mentioned there,
we can regard that Baker's theory is on the differentials of
the first kind over a hyperelliptic curve.
As compared with his theory, the ordinary soliton theories
{\it e.g.}, Sato theory [SS],
Date-Jimbo-Kashiwara-Miwa (DKJM) theory [DKJM],
Krichever theory [K1, K2], conformal field theory [KNTY]
and so on, can be considered as theories of
 the differentials of the second kind.
Thus Baker's theory is not directly connected with the modern
 soliton theories,
even though he used the Pfaffian, symmetric functions,
bilinear operator ${\bold{D}}$.
Indeed he might be interested only in
properties of  periodic functions on non-degenerate curves.
As long as I know,
 he did not consider the soliton
solutions, which is expressed by  hyperbolic functions or
trigonometric functions.
Hence he neither reached Hirota's direct method [H]
even though he defined and used the bilinear operator.

However, as all values appearing in Baker's theory have
algorithms to evaluate themselves, we can deal with
hyperelliptic functions in the framework of his theory
as we can do with elliptic functions.
For example, we can concretely determine any coefficients
of Laurent
or Taylor expansions of $\sigma$ and $\wp$ functions at any
points in any hyperelliptic curves [B1, B2, B3, G, \^O1, \^O3].
Recently requests to evaluate the hyperelliptic functions
explicitly appear from various fields, {\it e.g.},
 from study on the  Abel functions,
from number theory [G, \^O1, \^O3], and
from  study of  an elastica which is closely related to the
KdV equations [Ma1, Ma2].
There Baker's theory of
hyperelliptic functions plays a central role  [G, \^O1, \^O3, Ma2].
The purpose of this article is to reevaluate Baker's work
from the viewpoint of soliton theory.

\vskip 1.0 cm

After completion of this article, I knew the works of Buchstaber,
Enolskii and Leykin [BEL1-3] and others
[CEEK, EE, EEL, EEP, N and references therein].
The authors in [BEL1-3,CEEK, EE, EEL, EEP, N] also reevaluated
theory of Baker's hyperelliptic sigma functions, which they call
Kleinian functions, and have extended it from point of view of
soliton theory. For example in [B3], Baker derived a
differential identity of the hyperelliptic $\wp$-functions of
arbitrary genus, called  fundamental formula and mentioned in
\S4 of this article, which must include the KdV hierarchy and
the KP equations of higher genera but he explicitly presented them
only of genus three case. On the other hand, in [BEL1-2],
the authors  developed a method in terms of matrices by
considering a subset of $\wp$-functions
$(\wp_{gi})_{\{i=1,\cdots,g\}}$ as a vector and then gave the
explicit relation of the KdV hierarchy and the hyperelliptic
$\wp$-functions of arbitrary genus $g$. Their method is
consistent with the zero curvature condition in modern soliton
theory. Using the hyperelliptic sigma function and defining
natural sigma functions of more general algebraic curves, the
authors in [BEL1-3,CEEK, EE, EEL, EEP, N] have been
constructing deeper theories of abelian functions and soliton
equations. Thus it is needless to say that
[BEL1-3,CEEK, EE, EEL, EEP, N] are beyond the world of Baker.
In fact most of results in \S 2 of this article (proposition 4
and theorem 6) has been mentioned in their studies
[BEL1,2, EE] and review part of Baker's theory in [BEL2] is
very nice even for readers who are not familiar with
hyperelliptic functions. In [BEL3], it was pointed out that
$\wp_{11}$ of a hyperelliptic curve of genus $g>2$ with odd
degree polynomial is a solution of the KP equation, which
corresponds to the relation (IV-15) in (2-15) of this article.
However in [BEL1-3,CEEK, EE, EEL, EEP, N], they did not
comment upon the paper [B3], which  contains interesting and
fruitful results from modern point of view as described in \S4.
Further as far as I know, there has been no study on a
hyperelliptic function solution of the KP equation over a
hyperelliptic curve with even degree polynomial, which directly
reproduces the natural dispersion relations of the KP equation.
Connection between modern soliton theory [DKJM] and Baker's
theory discussed in \S 4 is also concerned from viewpoint of
the reevaluation. Thus I believe that this article is still
important.

\vskip 0.5 cm

{\centerline{\bf{\S 2. Hyperelliptic Solutions of KdV Equations}}}

\vskip 0.5 cm

In this section, we will consider hyperelliptic solutions of the first
and second KdV equations in the  KdV hierarchy.
First we will prepare notations and definitions on
this article. Although we mainly deal  with a curve
with genus three,
we give definitions and expressions of hyperelliptic curves
with general genus for later convenience.
In this article, we will mainly use the
conventions of \^Onishi [\^O1, \^O2].
We denote the set of complex number by $\Bbb C$ and
the set of integers by $\Bbb Z$.

\proclaim{\fp Notation 1}\it
We deal with a hyperelliptic curve $X_g$  of genus $g$
$(g>0)$ given by the algebraic equation,
$$ \split
   y^2 &= f(x) \\  &= \lambda_0 +\lambda_1 x
        +\lambda_2 x^2  +\cdots +\lambda_{2g+1} x^{2g+1}  \\
     &=(x-c_1)\cdots (x-c_g)(x-c_{g+1})
     \cdots (x-c_{2g})(x-c_{2g+1}),\\
\endsplit  \tag 2-1 $$
where $\lambda_{2g+1}\equiv1$ and $\lambda_j$'s and  $c_j$'s
are  complex values.

\endproclaim

Since we wish to treat the infinite point in this curve,
we should embed it in a projective space.
However as it is not difficult, we assume that
the curve $y^2=f(x)$ includes the infinite point.
Further for simplicity,
we also assume that $f(x)=0$ is not
degenerate.
We sometimes express a point $\mrm P$ in the curve
by the affine coordinate $(x,y)$.
\tvskip
\proclaim{\fp Definition 2 [B1 p.195, B2 p.314, B3
p.137, \^O1 p.385-6, \^O2]}\it

 \roster
\item  Let us denote the homology of a hyperelliptic
curve $X_g $ by
$$
\roman{H}_1(X_g, \Bbb Z)
  =\bigoplus_{j=1}^g\Bbb Z\alpha_{j}
   \oplus\bigoplus_{j=1}^g\Bbb Z\beta_{j},
 \tag 2-2
$$
where these intersections are given as
$[\alpha_i, \alpha_j]=0$, $[\beta_i, \beta_j]=0$ and
$[\alpha_i, \beta_j]=\delta_{i,j}$.

\item The unnormalized differentials of the first kind are defined by,
$$   \omega_1 := \frac{ d x}{2y}, \quad
      \omega_2 :=  \frac{x d x}{2y}, \quad \cdots, \quad
     \omega_g :=\frac{x^{g-1} d x}{2 y}.
      \tag 2-3
$$

\item The unnormalized differentials of the second kind are defined by,
$$
     \eta_{j}:=\dfrac{1}{2y}\sum_{k=j}^{2g-j}(k+1-j)
      \lambda_{k+1+j} x^k dx ,
     \quad (j=1, \cdots, g) .
     \tag 2-4
$$

\item The unnormalized period matrices are defined by,
$$    \pmb{\Omega}':=\left[\int_{\alpha_{j}}\omega_{i}\right],
\quad
      \pmb{\Omega}'':=\left[\int_{\beta_{j}}\omega_{i}\right],
 \quad
    \pmb{\Omega}:=\left[\matrix \pmb{\Omega}' \\ \pmb{\Omega}''
     \endmatrix\right].
  \tag 2-5
$$

\item The normalized period matrices are given by,
$$    \ ^t\left[\matrix \hw_{1}  \cdots & \hw_{g}
        \endmatrix\right]
       :={\pmb{\Omega}'}^{-1}  \ ^t\left[\matrix
          \omega_{1} & \cdots
   \omega_{g}\endmatrix\right] ,\quad
   \Bbb T:={\pmb{\Omega}'}^{-1}\pmb{\Omega}'',
   \quad
    \hat{\pmb{\Omega}}:=\left[\matrix 1_g \\ \Bbb T
     \endmatrix\right].
       \tag 2-6
$$

\item The complete hyperelliptic integrals of the second kind
are given  as
$$      H':=\left[\dint_{\alpha_{j}}\eta_{i}\right], \quad
         H'':=\left[\dint_{\beta_{j}}\eta_{i}\right] .
       \tag 2-7
$$

\item
By defining the Abel map for $g$-th symmetric product
of the curve $X_g$ and  for points $\{ Q_i\}_{i=1,\cdots,g}$
in the curve,
$$       \hat w: \roman{Sym}^g(X_g) \longrightarrow \Bbb C^g, \quad
      \left( \hat w_k(Q_i):=\sum_{i=1}^g \int_\infty^{Q_i}
      \hat \omega_k \right),
$$ $$ w:\roman{Sym}^g( X_g) \longrightarrow \Bbb C^g, \quad
     \left( w_k(Q_i):= \sum_{i=1}^g \int_\infty^{Q_i} \omega_k \right),
      \tag 2-8
$$
the Jacobi varieties $\hat{\Cal J_g}$ and $\Cal J_g$
are defined as complex torus,
$$   \hat{\Cal J_g} := \Bbb C^g /\hat{ \pmb{\Lambda}} ,
   \quad {\Cal J_g} := \Bbb C^g /{ \pmb{\Lambda}} .
     \tag 2-9
$$
Here $\hat{ \pmb{\Lambda}}$ $({ \pmb{\Lambda}})$  is a
lattice generated by
$\hat{\pmb{\Omega}}$ $({\pmb{\Omega}})$.

\item We defined the theta function over $\Bbb C^g$
characterized by $\hat{ \pmb{\Lambda}}$,
$$\theta\negthinspace\left[\matrix a \\ b \endmatrix\right](z)
    :=\theta\negthinspace\left[\matrix a \\ b \endmatrix\right]
     (z; \Bbb T)
    :=\sum_{n \in \Bbb Z^g} \exp \left[2\pi i\left\{
    \dfrac 12 \ ^t\negthinspace (n+a)\Bbb T(n+a)
    + \ ^t\negthinspace (n+a)(z+b)\right\}\right],
     \tag 2-10
$$
for $g$-dimensional vectors $a$ and $b$.

\endroster
\endproclaim

\tvskip
We should note that these contours in the integrals are,
for example,
given in p.3.83 in [M]. Thus above values can be, in principle,
computed in terms of numerical method for a given $y^2=f(x)$.

It is also noted that on (2-3), we have employed the
convention of \^Onishi [\^O1, \^O2], which differs
from Baker's original one by factor $1/2$. Due to the
difference, the results and definitions
in [B1, B2, B3] will be slightly
modified but the factor set us free from extra constant
factors in various situations [G, \^O1, \^O2, \^O3].

\tvskip

\vskip 0.5 cm
\proclaim {Definition 3 ($\wp$-function, Baker)
[B1, B2 p.336, p.358, p.370, \^O1 p.386-7, \^O2] }

\it
\roster
We prepare the coordinate in $\Bbb C^g$ for
 points $(x_i,y_i)_{i=1,\cdots,g}$
of the curve $y^2 = f(x)$,
$$
  \ft_j :=\sum_{i=1}^g\int^{(x_i,y_i)}_\infty \omega_j .
    \tag 2-11
$$

\item Using the coordinate $\ft_j$, sigma function,
which is a holomorphic
function over $\Bbb C^g$, is defined by
$$ \sigma(\ft)=\sigma(\ft;X_g):
  =\ \roman{exp}(-\dfrac{1}{2}\ ^t\ \ft
  H'{\pmb{\Omega}'}^{-1}\ft)
  \vartheta\negthinspace
  \left[\matrix \delta'' \\ \delta' \endmatrix\right]
  ({\pmb{\Omega}'}^{-1}\ft ;\Bbb T) .
     \tag 2-12
$$
where
$$
 \delta' =\ ^t\left[\matrix \dfrac {g}{2} & \dfrac{g-1}{2}
       & \cdots
      & \dfrac {1}{2}\endmatrix\right],
   \quad \delta''=\ ^t\left[\matrix \dfrac{1}{2} & \cdots
& \dfrac{1}{2}
   \endmatrix\right].
     \tag 2-13
$$

\item
In terms of $\sigma$ function, $\wp$-function over the
hyperelliptic curve is given by
$$   \wp_{\mu\nu}(\ft)=-\dfrac{\partial^2}{\partial
   \ft_\mu\partial \ft_\nu}
   \log \sigma(u) .
         \tag 2-14
$$
\endroster

\endproclaim

The $\sigma$-function is a well-tuned theta-function.
(2-13) is related to so-called Riemannian constant $K$ as mentioned
in p.3.80-82 in [M];
$\delta'+\Bbb T\delta''$ agrees with $K$.
 As the $\sigma$-function [B2, p.336, p.358]
consists of the shifting Riemann theta function (2-10)
[B2, p.324, p.336], the Riemann constant $K$ outwardly
disappears. (Thus the $\sigma$-function vanishes just over the
theta divisor.)
Using the $\sigma$-function, Baker derived the multiple relations of
$\wp$-functions and so on. Hereafter we assume that genus of
the curve is three.

\proclaim{\fp Proposition 4 [B3 p.155-6,\^O1 p.388,\^O2]}\it

Let us express
$\wp_{\mu\nu\rho}:=\partial \wp_{\mu\nu}(\ft)
  /\partial \ft_\rho$ and
$\wp_{\mu\nu\rho\lambda}:=\partial^2
 \wp_{\mu\nu}(\ft) /\partial \ft_\mu \partial \ft_\nu$.
Then hyperelliptic $\wp$-functions obey the relations,
$$   \allowdisplaybreaks \align
     ( \IV-1)\quad& \wp_{3333}-6\wp_{33}^2
      =  2\lambda_5 \lambda_7
       + 4\lambda_6 \wp_{33} + 4 \lambda_7 \wp_{32},\\
     ( \IV-2)\quad& \wp_{3332}-6\wp_{33} \wp_{32}
      =  4\lambda_6 \wp_{32} + 2 \lambda_7 (3\wp_{31}
        -\wp_{22}),\\
     ( \IV-3)\quad& \wp_{3331}-6\wp_{31}\wp_{33}
      =  4\lambda_6 \wp_{31} - 2 \lambda_7 \wp_{21},\\
     ( \IV-4)\quad& \wp_{3322}-4\wp_{32}^2-2\wp_{33} \wp_{22}
      =  2\lambda_5 \wp_{32} + 4 \lambda_6 \wp_{31} - 2
       \lambda_7 \wp_{21},\\
     ( \IV-5)\quad& \wp_{3321}-2\wp_{33} \wp_{21}-4\wp_{32} \wp_{31}
      =  2\lambda_5 \wp_{31},\\
     ( \IV-6)\quad& \wp_{3311}-4\wp_{31}^2-2\wp_{33} \wp_{11}
      = 2 \Delta, \\
     ( \IV-7)\quad& \wp_{3222}-6\wp_{32}\wp_{22}
      = -4\lambda_2 \lambda_7
        -2\lambda_3 \wp_{33} + 4 \lambda_4 \wp_{32}
        +4\lambda_5 \wp_{31} - 6 \lambda_7 \wp_{11},\\
     ( \IV-8)\quad& \wp_{3221}-4\wp_{32} \wp_{21}-2\wp_{31} \wp_{22}
      =- 2\lambda_1 \lambda_7+ 4 \lambda_4 \wp_{31} - 2  \Delta,
          \\
     ( \IV-9)\quad& \wp_{3211}-4\wp_{31} \wp_{21}-2\wp_{32} \wp_{11}
      =- 4\lambda_0 \lambda_7+ 2 \lambda_3 \wp_{31},\\
     (\IV-10)\quad& \wp_{3111}-6\wp_{31} \wp_{11}
      =  4\lambda_0 \wp_{33} - 2 \lambda_1 \wp_{32}
        + 4 \lambda_2 \wp_{31}, \\
     (\IV-11)\quad& \wp_{2222}-6\wp_{22}^2
         =-8\lambda_2 \lambda_6+ 2 \lambda_3 \lambda_5\\
        &- 6 \lambda_1 \lambda_7
       -12\lambda_2 \wp_{33} + 4 \lambda_3 \wp_{32}
           + 4 \lambda_4 \wp_{22}
       + 4\lambda_5 \wp_{21} -12 \lambda_6 \wp_{11}
         +12 \Delta,\\
     (\IV-12)\quad& \wp_{2221}-6\wp_{22} \wp_{21}
      =- 4\lambda_1 \lambda_6- 8 \lambda_0 \lambda_7
       - 6\lambda_1 \wp_{33} + 4 \lambda_3 \wp_{31}
          + 4 \lambda_4 \wp_{21}
       - 2\lambda_5 \wp_{11}, \\
     (\IV-13)\quad& \wp_{2211}-4\wp_{21}^2
            -2\wp_{22} \wp_{11}
      =- 8\lambda_0 \lambda_6
       - 8\lambda_0 \wp_{33} - 2 \lambda_1 \wp_{32}
         + 4 \lambda_2 \wp_{31}
       + 2\lambda_3 \wp_{21}, \\
     (\IV-14)\quad& \wp_{2111}-6\wp_{21} \wp_{11}
      =- 2\lambda_0 \lambda_5
       - 8\lambda_0 \wp_{32} + 2 \lambda_1(3\wp_{31} -\wp_{22})
       + 4\lambda_2 \wp_{21}, \\
     (\IV-15)\quad& \wp_{1111}-6\wp_{11}^2
      =- 4\lambda_0 \lambda_4+ 2 \lambda_1 \lambda_3
       + 4\lambda_0(4\wp_{31}- 3 \wp_{22})
       + 4\lambda_1 \wp_{21} + 4 \lambda_2 \wp_{11},
       \tag 2-15
     \endalign
$$ where $$
     \Delta  = \wp_{32} \wp_{21} - \wp_{31} \wp_{22}
     + \wp_{31}^2 - \wp_{33} \wp_{11}.
      \tag 2-16
$$

\endproclaim

\proclaim{\fp Remark 5}\it
\roster

\item
Due to the definitions, indices of $\wp$ are symmetric,
{\it i.e.}, $\wp_{\mu\nu}=\wp_{\nu\mu}$,
$\wp_{\mu\nu\rho}=\wp_{\rho\mu\nu}=\wp_{\nu\rho\mu}$
and so on.

\item
Above equations are independent because the axes of Jacobian
$\Cal J_g$ are independent.

\item
As Baker did [B3, p.151],
introducing  the bilinear differential operator ${\bold{D}}_\nu$,
$$
        {\bold{D}}_\mu \sigma(u) \sigma(u):=
                    (\frac{\partial}{\partial u'_\mu}
    - \frac{\partial}{\partial u_\mu}) \sigma(u')
       \sigma(u)|_{u=u'},
      \tag 2-17
$$
we have the relations,
$$
        \wp_{\mu\nu} =-\frac{1}{2 \sigma^2}
   {\bold{D}}_\mu {\bold{D}}_\nu \sigma \sigma,
     \tag 2-18
$$
$$
        \wp_{\lambda\mu\nu\rho}-2(\wp_{\mu\nu}\wp_{\lambda\rho}
       +\wp_{\nu\lambda}\wp_{\rho\mu}
       +\wp_{\lambda\mu}\wp_{\rho\nu})
 =-\frac{1}{2 \sigma^2}{\bold{D}}_\lambda
   {\bold{D}}_\mu {\bold{D}}_\nu {\bold{D}}_\rho \sigma \sigma.
       \tag 2-19
$$
Then
the equations in Proposition 4
can be regarded as the bilinear equations
of $\sigma$-functions. For example, (IV-1) is given by
$$
  ({\bold{D}}_3^4-4 \lambda_6 {\bold{D}}_3^2 -4 {\bold{D}}_3
      {\bold{D}}_2
     - 4 \lambda_5 \lambda_7 ) \sigma\sigma =0 .
      \tag 2-20
$$

 \endroster
\endproclaim

\proclaim{\fp Theorem 6}\it

For $v=-2(\wp_{33}+\lambda_{6}/3 )$ and
$v(t_1,t_3,t_5)=v(\ft_3, -\dfrac{\ft_{2}}{2^2},
\dfrac{\ft_{1}}{2^4} + \dfrac{3}{2^4\lambda_{6}} \ft_{2})$
obeys first and second KdV equations:
$$
        \partial_{t_3} v + 6 v \partial_{t_1} v
        +\partial_{t_1}^3 v=0,
       \tag 2-21
$$
$$
\partial_{t_5} v + 30 v^2 \partial_{t_1} v
+20\partial_{t_1} v\partial_{t_1}^2 v+
10v \partial_{t_1}^3 v+\partial_{t_1}^5 v=0.
    \tag 2-22
$$

\endproclaim

\demo {Proof} By differentiating (IV-1) in $u_3$ and tuning them,
we obtain the KdV equation. We note that second KdV equation
is expressed by
$$
\partial_{t_5} v +( \partial_{t_1}^2 + 2 v
+ 2 \partial_{t_1} v \partial_{t_1}^{-1}) (6 v \partial_{t_1}v
  +
  \partial_{t_1}^3 v )=0,
 \tag 2-23
$$
where $\partial_{t_1}^{-1}$ implies an integral with respect to
$t_1$.
By setting $2 \partial_{\ft_3}
\times (\IV-2) + \partial_{\ft_2}\times (\IV-1)$ and $\partial_{t_5}=
16\partial_{\ft_1}+ \dfrac{16\lambda_2}{3} \partial_{\ft_2}$,
we obtain second KdV equation.\qed \enddemo

\proclaim{\fp Remark 7}\it
\roster

\item Theorem 6 and definition of $\wp$ mean that solutions of
the KdV equation are explicitly constructed. The quantities in
definitions 2 and 3 can be, in principle,
evaluated in terms of numerical computations
because there is no ambiguous parameter.

\item
We note the dispersion relations:
$\ft_j$ behaves like $(1/\bx)^{2(g-j)+1}$ around
$\infty$ point if we use local coordinate $\bx^2 :=x$. By
comparing the order of $\bx$ denoted by $\ord_{\bx}$,
we have the relations,
$$
        \ord_{\bx}(\ft_2) = 3 \ord_{\bx}(\ft_3),
\quad \ord_{\bx}(\ft_1)=5\ord_{\bx}(\ft_3).
         \tag 2-24
$$
These are the dispersion relations of the KdV
equations.

\item
Roughly speaking
integrating the KdV equation in $t_1$ becomes (\IV-1)
in proposition 4.
Then there appears an undetermined integral constant.
However in proposition 4, it is fixed
and associated with the coefficients of the algebraic
equation $y^2=f(x)$. Thus
 (IV-1) in proposition 4 is more fundamental
than the KdV equation.

\item For genus two case: we put that
$\partial \sigma/\partial u_3=0$
and $\lambda_6=\lambda_7=0$; (IV-1)-(IV-10) becomes meaningless $0=0$
and $\Delta=0$. $v=-2(\wp_{22}+\lambda_{4}/3 )$ and
$v(t_1,t_3)=v(\ft_2, -\dfrac{\ft_{1}}{2^2})$
obeys first  KdV equation (2-21).

\item For genus one case or elliptic functions case:
we put that $\partial \sigma/\partial u_\mu=0$ ($\mu=2,3$),
and $\lambda_a=0$ $(a=4,5,6,7)$; only (IV-15) survives,
which is the relation of elliptic $\wp$ function.
\endroster

\endproclaim

\vskip 0.5 cm

{\centerline{\bf{\S 3. Hyperelliptic Solutions of KP Equation}}}

\vskip 0.5 cm
Instead of the curve of $(2g+1)$-degree, we will deal with a
hyperelliptic curve of $(2g+2)$-degree in this section.

\proclaim{\fp Notation 8}\it
$$ \split
   y^2 &= \bbf(x) \\  &= \blambda_0 +\blambda_1 x
        +\bar\lambda_2 x^2  +\cdots +
         \bar\lambda_{2g+2} x^{2g+2}  \\
     &=(x-\alpha_1)\cdots (x-\alpha_g)(x-\alpha_{g+1})
     \cdots (x-\alpha_{2g})(x-\alpha_{2g+1})(x-\alpha_{2g+2}),
\endsplit
\tag 3-1
$$
where $\bar\lambda_{2g+2}\equiv1$ and $\bar\lambda_j$'s
and  $\alpha_j$'s are complex values.

\endproclaim

\proclaim{\fp Remark 9 [B1 p.195, B3 p.144-5]}\it
\roster

\item
The transformation  between $y^2=f(x)$ and $\zeta^2 = \bbf(\xi)$
is as follows
$$
  x=\frac{a}{\xi-\alpha_{2g+2}}, \quad
 c_i = \frac{a}{\alpha_i-\alpha_{2g+2}}, \quad
\zeta = \frac{(\xi-\alpha_{2g+2})^{g+1}}{
-4\prod_i^{2g-1} c_j} y.
\tag 3-2
$$

\item The unnormalized differentials of the first kind are defined by,
$$   \omega_1 = \frac{ d x}{2y}, \quad
      \omega_2 =  \frac{x d x}{2y}, \quad \cdots, \quad
     \omega_g =\frac{x^{g-1} d x}{2 y}.
\tag 3-3
$$

\item The unnormalized differentials of the second kind are defined by
([B2, p.195]),
$$
     \eta_{j}=\dfrac{1}{2y}\sum_{k=j}^{2g+1-j}(k+1-j)
      \bar\lambda_{k+1+j} x^k dx ,
     \quad (j=1, \cdots, g) .
\tag 3-4
$$

\item
The definition of $\sigma$ and $\wp$-functions
 are the same as those in definition 2
and 3, where we regard that $y$  obeys the equation
$y^2=\bbf(x)$ instead of $y^2=f(x)$.

\endroster
\endproclaim

\proclaim{\fp Proposition 10   [B3 p.155-6]}\it

The hyperelliptic $\wp$-functions of a curve
$y^2 = \bbf(x)$ ($g=3$)
obey the relations
$$   \allowdisplaybreaks \align
     ( \X-1)\quad& \wp_{3333}-6\wp_{33}^2
      =  2\blambda_5 \blambda_7
       + 4\blambda_6 \wp_{33} + 4 \blambda_7 \wp_{32}
        - 8 \blambda_4 \blambda_8 + 4 \blambda_8(4 \wp_{31}
        -3 \wp_{22}),\\
     ( \X-2)\quad& \wp_{3332}-6\wp_{33} \wp_{32}
      =  4\blambda_6 \wp_{32} + 2 \blambda_7 (3\wp_{31}-\wp_{22})
        -4\blambda_3 \blambda_8 + 8 \blambda_8 \wp_{21},\\
     ( \X-3)\quad& \wp_{3331}-6\wp_{31}\wp_{33}
      =  4\blambda_6 \wp_{31} - 2 \blambda_7 \wp_{21}+4
        \blambda_8 \wp_{11},\\
     ( \X-4)\quad& \wp_{3322}-4\wp_{32}^2-2\wp_{33} \wp_{22}
      =  2\blambda_5 \wp_{32} + 4 \blambda_6 \wp_{31} - 2
         \blambda_7 \wp_{21}
        -8\blambda_2\blambda_8 - 8 \blambda_8 \wp_{11},\\
     ( \X-5)\quad& \wp_{3321}-2\wp_{33} \wp_{21}-4\wp_{32} \wp_{31}
      =  2\blambda_5 \wp_{31}-4\blambda_1\blambda_8,\\
     ( \X-6)\quad& \wp_{3311}-4\wp_{31}^2-2\wp_{33} \wp_{11}
      = 2 \Delta, \\
     ( \X-7)\quad& \wp_{3222}-6\wp_{32}\wp_{22}
      = -4\blambda_2 \blambda_7
        -2\blambda_3 \wp_{33} + 4 \blambda_4 \wp_{32}
        +4\blambda_5 \wp_{31} - 6 \blambda_7 \wp_{11}-8\blambda_1
          \blambda_8,\\
     ( \X-8)\quad& \wp_{3221}-4\wp_{32} \wp_{21}-2\wp_{31} \wp_{22}
      =- 2\blambda_1 \blambda_7+ 4 \blambda_4 \wp_{31}
       - 2  \Delta-
         8\blambda_0\blambda_8,
     \tag 3-5 \endalign
$$
 (9)-(15) and $\Delta$ which have  the same form as those
in proposition 4
by replacing $\lambda$'s with $\blambda$'s.

\endproclaim

\proclaim{\fp Theorem 11}\it

For $v=-2(\wp_{33}+\blambda_{6}/3 )$ and
$u(t_1,t_2,t_3)=v(\ft_3, \dfrac{\ft_{2}}{2\sqrt{-3}},
-\dfrac{\ft_{1}}{2^4}
- \dfrac{3}{2^2\blambda_{7}} \ft_{2})$
obeys the KP equation:
$$
        \partial_{t_1}(\partial_{t_3} v
          + 6 v \partial_{t_1} v+\partial_{t_1}^3 v)
         =\partial_{t_2}^2 v.
\tag 3-6
$$

\endproclaim
\demo {Proof} Noting $\blambda_8=1$, direct substitution of them into
(3-6) is differential of (X-1) in $u_3$. \qed \enddemo

\proclaim{\fp Remark 12}\it

\roster
\item
Theorem 11 means that we obtain an explicit function form
of hyperelliptic function solution of the KP equation.

\item
We note the dispersion relation.
Since the curve $y^2 =\bbf(x)$ is not ramified
at infinity point. There $\ft_j$ behaves like $(1/x)^{(g-j)}$
upto a constant factor.
By
comparing the order of $x$ denoted by $\ord_{x}$,
we have the relations,
$$
        \ord_{x}(\ft_2) = 2 \ord_{x}(\ft_3),
\quad \ord_{x}(\ft_1)=3\ord_{x}(\ft_3).
\tag 3-7
$$
These are the dispersion relations of the KP
equation.
\endroster
\endproclaim
\vskip 0.5 cm

{\centerline{\bf{\S 4.  Discussion}}}

\vskip 0.5 cm
Since derivation of proposition 10 is essentially the same
as that of proposition 4,
 we will give a sketch only of the derivation  of
the differential equations in proposition 4 and comment upon
its relation to the soliton theory.

 \definition{\bf Definition 13 [B1 p.195, B2 p.314, p.335-6, \^O2]}

\it
{\roster
For points of $\mrm P(x,y)$, $\mrm Q(z,w)$,
$\mrm A(a,b)$, $\mrm B(c,d)
$ over $X_g$, we introduce the quantities,

\item
 $$\bold{ R}_{\mrm Q,\mrm B}^{\mrm P,\mrm A}:=
    \dint_{\mrm A}^{\mrm P}\dint_{\mrm B}^{\mrm Q}
    \dfrac{f(x,z)+2yw}{(x-z)^2}\dfrac{dx}{2y}\dfrac{dz}{2w},
\tag 4-1
$$
where
$$
f(x,z):=\sum_{j=0}^{g}x^jz^j(\lambda_{2j+1}(x+z)+2\lambda_{2j}).
\tag 4-2
$$

\item
  $$
  \bold{ P}_{\mrm Q, \mrm B}^{\mrm P, \mrm A}
  :=\int_{\mrm A}^{\mrm P}\left(\frac{y+w}{x-z}
              -\frac{y+d}{x-c}\right)\frac{dx}{2y} .
  \tag 4-3
  $$

\endroster

\enddefinition

\proclaim{\fp Proposition 14 [B1 p.194-5, B2 p.318, p.336, \^O2]}
\it
\roster
\item ${\bold R}_{\mrm Q,\mrm B}^{\mrm P,\mrm A}$ and
$\bold{ P}_{\mrm Q, \mrm B}^{\mrm P, \mrm A}$ as
a function of $\mrm P$ have singularity around
$\mrm P=\mrm Q$, $\mrm B$ of first order with the
residues $1$, $-1$ and holomorphic otherwise.
In other words, they are unnormalized third differentials.

\item
  $$
  \bold{R}_{\mrm Q,\mrm B}^{\mrm P,\mrm A}
  =\int_{\mrm A}^{\mrm P}\w_{1}\int_{\mrm B}^{\mrm Q}\eta_{1}
   +\cdots
  +\int_{\mrm A}^{\mrm P}\w_{g}\int_{\mrm B}^{\mrm Q}\eta_{g}
  +\bold{P}_{\mrm Q, \mrm B}^{\mrm P, \mrm A} .
   \tag 4-4
  $$

\item
For ${\mrm P_j}$, ${\mrm Q_j} \in X$, ($j=1$, $\cdots$, $g$),
and
$$
  u=\sum_{j=1}^g\int_{\infty}^{\mrm P_j}\omega, \quad
  u'=\sum_{j=1}^g\int_{\infty}^{\mrm Q_j}\omega,
   \tag 4-5
$$
the following relation holds,
  $$
  \exp\left(\sum_{j=1}^g\bold{ R}
  _{\overline{\mrm P_j},
      \overline{\mrm Q_j}}^{\mrm P, \mrm Q}\right)
  =\dfrac{\sigma\left(\dint_{\infty}^{\mrm P}\omega+u\right)
          \sigma\left(\dint_{\infty}^{\mrm Q}\omega+u'\right)}
         {\sigma\left(\dint_{\infty}^{\mrm P}\omega+u'\right)
          \sigma\left(\dint_{\infty}^{\mrm Q}\omega+u\right)} ,
       \tag 4-6
  $$
where $\overline{\mrm P_j}$ $(\overline{\mrm Q_j})$
is conjugate of ${\mrm P_j}$ $({\mrm Q_j})$
with respect to the symmetry of hyperelliptic curve
$(x,y) \to (x,-y)$.
\endroster
\endproclaim

\proclaim{\fp Remark 15}

The relation 4-6 is very important. It holds for
appropriate $\sigma$-functions and  third
differentials in a general
compact Riemannian surface [B1 p.290], even though their form
can not globally written like definition 13. As we show below,
the relation plays important roles in both Baker's theory
and DKJM-theory [DKJM].

\endproclaim

Here we will sketch the derivation of the equations
in propositions 4 following [B1] and [B3].
First we introduce the variables for the divisors
${\mrm P_j}=(x_j, y_j)$
and ${\mrm P}=(x, y)\equiv(x_0,y_0)$ in notations
in proposition 14 (3),
$$
        \frak t:=(\int^P_\infty \omega+u), \tag 4-7
$$
$$
        R(z):=(z-x_0)F(z):=(z-x_0)(z-x_1)(z-x_2)\cdots(z-x_g),
        \tag 4-8
$$
$$
        \frac{R(z)}{(z-x_r)(z-x_s)}=:
           z^{g-1} + c_1^{r,s} z^{g-2}+ c_2^{r,s} z^{g-3}
          +\cdots+c_g^{r,s},
          \tag 4-9
$$
and for generic parameter $e$,
$$
 \overline \delta_{e}:=\sum_{\mu=1}^g e^{\mu-1}
        \frac{\partial}{\partial \frak t_\mu}.
       \tag 4-10
$$
We operate $ \overline\delta_{e_1}\overline\delta_{e_2}$ to
the both sides in the relation (4-6) in proposition 14.
We should note the relation,
$$
        \sum_{r=0, r\neq s}^g
       \frac{x_r-x_s}{R'(x_r)}c_{l-1}^{r,s}
         x_r^{g-k} = \delta_l^k,
         \tag 4-11
$$
where $c^{r,s}_0=1$ and $R'(x_r) = d R(z)/d z|_{z=x_r}$.
By taking limit  $x_0 \to \infty$, we obtain [B1 p.328, p.376]
$$
\split
(e_1-e_2)^2&\sum_{\lambda=1}^g \sum_{\mu=1}^g
  \wp_{\lambda\mu}(u)
 e_1^{\lambda-1} e_2^{\mu-1}\\
&=\left(\sum_{r=1,s=1}^g
\frac{F(e_1)F(e_2)(2y_ry_s-f(x_r,x_s))}
  {(e_1-x_r)(e_2-x_r)(e_1-x_s)(e_2-x_s) F'(x_r)F'(x_s)}\right).
\endsplit
        \tag 4-12
$$
We deform it by shifting the zero of $\wp$ to obtain
[B1, p.328, B3 p.138],
$$
\split
\sum_{\lambda=1}^g \sum_{\mu=1}^g \wp_{\lambda\mu}(u)
 e_1^{\lambda-1} e_2^{\mu-1}&=
F(e_1)F(e_2)\left(\sum_{r=1}^g
\frac{y_r}
  {(e_1-x_r)(e_2-x_r)F'(x_r)}\right)^2\\
&-\frac{f(e_1)F(e_2)}
  {(e_1-e_2)^2F'(e_1)}-\frac{f(e_2)F(e_1)}
  {(e_1-e_2)^2F'(e_2)}+\frac{f(e_1,e_2)}{(e_1-e_2)^2}.
\endsplit
       \tag 4-13
$$
Even though in [B3] Baker adopted this formula (4-13)
 as a
definition of $\wp$-functions,
his arguments on this formula stood upon the background
of so many studies on
the hyperelliptic function [B1, B2].
Thus we should regard (4-13) as a theorem which was
proved in [B1].

Introducing another operator,
$$
\delta_{e}=\frac{1}{F(e)}\sum_{j=1}^g e^{j-1}
\frac{\partial}{\partial u_j},
 \tag 4-14
$$
we operate $\delta_{e_3}\delta_{e_4}$ to above relation (4-13)
and then we have "fundamental formula" [B3, p.144].
The section I in [B3] devoted the derivation of his fundamental
formula, which is very
tedious and complex but somewhat attractive. In fact,
tracing his derivations makes me feel that there might be
deep symmetry behind his theory.
In section II in [B3], Baker concentrated genus three case.
By comparing the coefficients of each $e_1^a e_2^b e_3^c e_4^d$,
he discovered the differential equations in propositions 4 and 10.
In the comparison, Baker used the symmetric functions,
Pfaffian and bilinear operators.
The symmetric functions naturally appears because
the differential of the first kind
 in the hyperelliptic curve is expressed by [B3],
$$
   \pmatrix d u_1 \\ d u_2 \\ du_3 \\ . \\ .\\ d u_g\endpmatrix
   =\frac{1}{2}\pmatrix 1/y_1      & 1/y_2     & \cdots & 1/y_g \\
            x_1/y_1    & x_2/y_2   & \cdots & x_g/y_g \\
           x_1^2/y_1    & x_2^2/y_2   & \cdots & x_g^2/y_g \\
       .            & .          & \cdots & .         \\
               . & .          & \cdots & .         \\
            x_1^{g-1}/y_1    & x_2^{g-1}/y_2
         & \cdots & x_g^{g-1}/y_g
        \endpmatrix
        \pmatrix d x_1 \\ d x_2 \\ d x_3 \\ .\\ .\\ d x_g
        \endpmatrix. \tag 4-15
$$
This matrix is resemble to Vandermonde matrix. In fact (4-11)
is an identity used in construction of inverse matrix of
Vandermonde matrix.

Corresponding to the above matrix (4-15),
behavior of
differentials of the second kind in theory of KP hierarchy
[K1, K2, SS, DKJM, KNTY] is sometimes
determined by the Vandermonde matrix,
$$
        \pmatrix 1      & 1   & \cdots & 1 \\
                  \bx_1    & \bx_2   & \cdots & \bx_p \\
                  \bx_1^2    & \bx_2^2  & \cdots & \bx_p^2\\
        .            & .          & \cdots & .         \\
      \bx_1^{p-1}    & \bx_2^{p-1}   & \cdots & \bx_p^{p-1}
       \endpmatrix. \tag 4-16
$$
The difference between Baker's theory of hyperelliptic function
and modern soliton theory could be regarded as the difference
between (4-15) and (4-16).

In modern soliton theory [SS, DKJM, KNTY],
we deal with a formal graded ring
$G\Bbb C[[\bx]]$ $:=\cup_n G^n\Bbb C[[\bx]]$ related
to degrees of $\bx$'s as a localized ring at infinity point
of an algebraic curve.
Then we
consider maps among  quotient modules
$ G^n\Bbb C[[\bx]]$ $/G^{n-1}\Bbb C[[\bx]]$, which consists of
 $\partial_{\bx}$ and $\bx$.
The differential ring generated by $\partial_{\bx}$ and $\bx$
 becomes Sato theory [SS] and conformal field theory
[KNTY]
after appropriately modifying it.
There naturally appear the Vandermonde matrix (4-16) of $\bx$'s,
symmetric functions, Pfaffian related to
behavior of differential of the second kind around the infinity point;
the Vandermonde determinate is related to Fermion amplitude
[DKJM, KNTY].

In the theory of differentials of the second kind, when one determines
the global behavior of algebraic function on a curve
by its local data around infinity point, he uses
the properties of holomorphic functions over the curves,
such as existence theorem, flabby of related sheaves and so on.
On the other hand, Baker's theory is of  differentials of 
the first kind
and  it is a global theory because differentials of the first kind
are holomorphic allover the curve and explicitly given.
 Accordingly we can deal with the
hyperelliptic functions in the framework of
Baker's theory as we do with elliptic functions.

\vskip 0.5 cm

We will comment on the proposition 4 in the framework of DKJM-theory
[DKJM].
\proclaim{Remark 16}

{\it\roster

For  points $P=(x,y)$, $Q=(\sqrt{-1}x,y)$ and
$\overline{P}=(x,-y)$
around the infinity points $x=\bx^2$, we obtain the following
relations

\item
$$
\bold{ R}
  _{\overline{\mrm A}, \overline{\mrm B}}^{\mrm P, \mrm Q}=
\bold{ R}
  ^{\overline{\mrm A}, \overline{\mrm B}}_{\mrm P, \mrm Q}
      \tag 4-17
$$

\item
$$
\split
        \int^{(x,y)}_{\infty} \omega_\mu
        & =-\int^{(\sqrt{-1}x,y)}_{\infty} \omega_\mu
     = \int^{(x,y)}_{(\sqrt{-1}x,y)} \frac{x^{\mu-1}d x}{y}\\
            &= -\frac{1}{2g-2\mu+1} \frac{1}{\bx^{2g-2\mu+1}}+
              \text{\rm{ lower oder terms}}.
\endsplit
         \tag 4-18
$$

\item
$$
        \int^{(x,y)}_{(\sqrt{-1}x,y)} \eta_j =2[\bx^{2g-2j+1}]+
\text{\rm{ lower oder terms}}.
          \tag 4-19
$$

\item
$$
\split
\sum_{j=1}^g\bold{ R}_{\overline{\mrm P_j},
              \overline{\mrm Q_j}}^{\mrm P, \mrm Q}
       =&-2[(u_1-u_1')\bx^{2g-1}+(u_2-u_2')\bx^{2g-3}+\cdots
        +(u_g-u_g')\bx]
        +\sum_{j=1}^g\bold{ P}_{\overline{\mrm P_j},
              \overline{\mrm Q_j}}^{\mrm P, \mrm Q}\\
&+\text{\rm{ lower oder terms}}.
\endsplit
         \tag 4-20
$$

\endroster}

\endproclaim

Using the remark 16 and setting $g=\infty$,
the relation (3) in proposition 14 is reduced to
the generating relation of the KdV hierarchy in DKJM-method:
$$
\split
\oint_\infty \frac{d \bx}{\bx}
        &\exp(\sum_{j=1}^g(u_j-u_j')\bx^{2g-2i+1})
        \sigma(u_1-\frac{1}{2g-1}\frac{1}{\bx^{2g-1}},
                        u_2-\frac{1}{2g-3}\frac{1}{\bx^{2g-3}},
         \cdots ,u_g-\frac{1}{\bx})\\
 &\quad
        \sigma(u_1'+\frac{1}{2g-1}\frac{1}{\bx^{2g-1}},
                        u_2'+\frac{1}{2g-3}\frac{1}{\bx^{2g-3}},
         \cdots ,u_g'+ \frac{1}{\bx})
       =0.
\endsplit
         \tag 4-21
$$
In terms of  differential operators, we can rewrite this relation
and then we obtain the KdV hierarchy [DKJM].
Thus the origins of the KdV hierarchy in Baker's method and
DKJM-method are the same.

\proclaim{Remark 17}

We will summary the difference between the soliton theory and
Baker's theory.

\roster

\item
As in soliton-theory of the KdV hierarchy [DKJM, K1, K2, SS],
 we investigate the behavior of
meromorophic functions around infinity point of
a hyperelliptic curve,
1-1) it can be regarded as a theory of differentials of
the second kind,
1-2) it can be extended to theory of meromorophic
functions of a general compact Riemannian surface as
the theory of the KP hierarchy [DKJM, K1, K2, KNTY, SS],
 and 1-3) we can not determine fine structure of meromorophic
functions of non-degenerate curve.

\item As in the Baker's theory of hyperelliptic $\wp$-functions,
we consider behavior of $\wp$-functions around
generic points $(x_1,y_1)$, $\cdots$ $(x_g,y_g)$ of a
hyperelliptic curve,
2-1) we directly deal with differentials of first kind which
are holomorphic allover the curve,
2-2) we can determine all parameters in $\wp$-functions of
the curve,
2-3) we can give explicit function forms of $\wp$-functions and
coefficients of Laurent expansions around any points in the
curve,
and 2-4) we can not extend it to general compact Riemannian
surface with this concreteness.

\item The differentials of the first kind and the second kinds
are complementarily connected as the term in (4-4) of
the most important identity (4-6). Thus in (4-6),
they behaves like two sides of the same coin.

\endroster
\endproclaim

Finally we comment upon this study.
In Baker's theory, we have no ambiguous and
dependent parameters
while in ordinary soliton theory of periodic solutions
there appear
undetermined parameters which must satisfy several relations.
Hirota and Ito gave explicit function forms of hyperelliptic
functions of genera two and three as periodic solutions
of the KdV equation (2-21) [HI]; they determined several parameters
by means of numerical computations. However functions
should be expressed only by independent variables and
thus Baker's theory has the advantage and is appropriate
even from viewpoint of numerical study.
I hope that in near future,
anyone would be able to
 plot graph of any
hyperelliptic functions or any periodic multi-soliton solutions
like graphs in [HI],
using a personal computer and a laser printer,
as we can do for elliptic functions or elliptic soliton
solutions.

{\centerline{\bf{ Acknowledgment}}}

I'm deeply indebted to  Prof.~Y.~\^Onishi
for leading me this beautiful theory
of Baker and to Prof.~K. Tamano and
H.~Mitsuhashi for fruitful discussions.
I thank Prof. V. Z. Enolskii and Prof. R. Hirota for sending me
their interesting works.
I am also grateful to both referees for helpful comments.

\enddocument